\documentclass[letterpaper,10 pt,conference]{ieeeconf} 
\IEEEoverridecommandlockouts 
\usepackage{graphicx}
\usepackage{euscript,color}
\usepackage{amssymb,amsfonts,amsmath,amscd,dsfont,mathrsfs}
\usepackage{cite}

\DeclareMathOperator*{\argmin}{arg\,min}

\usepackage{xspace}
\usepackage{bbm}

\newcommand{\Ac}{\mathcal{A}}
\newcommand{\Bc}{\mathcal{B}}

\newcommand{\Ec}{\mathcal{E}}

\newcommand{\Nc}{\mathcal{N}}

\newcommand{\Sc}{\mathcal{S}}

\newcommand{\Uc}{\mathcal{U}}

\newcommand{\xv}{{\bf x}}

\newcommand{\xh}{{\hat{x}}}

\newcommand{\eh}{{\hat{e}}}

\def\g{\gamma}

\def\e{\epsilon}

\def\t{\theta}

\let\P\relax
\DeclareMathOperator\P{P}

\newcommand\ie{i.e.,\xspace}
\def\textiid{i.i.d.\@\xspace}
\newcommand\iid{\ifmmode\text{ i.i.d. } \else \textiid \fi}

\newcommand{\ind}{\mathbbmss{1}}

\newcommand{\x}{{\bf x}}

\newtheorem{definition}{Definition}

\newtheorem{theorem}{Theorem}
\newtheorem{lemma}{Lemma}

\newtheorem{corollary}{Corollary}

\newtheorem{proposition}{Proposition}

\begin{document}

\title{\LARGE \bf Minimum Complexity Pursuit: Stability Analysis}

\author{\authorblockN{Shirin Jalali}
\authorblockA{Center for Mathematics of Information\\
California Institute of Technology\\
Pasadena, California, 91125}
\and
\authorblockN{Arian Maleki}
\authorblockA{Department of Electrical Engineering\\
Rice University\\
Houston, Texas, 77005} 
\and
\authorblockN{Richard Baraniuk}
\authorblockA{Department of Electrical Engineering\\
Rice University\\
Houston, Texas, 77005}
}

\maketitle

\newcommand{\p}{\mathds{P}}
\newcommand{\mb}{\mathbf{m}}   
\newcommand{\bb}{\mathbf{b}}

\begin{abstract}

A host of problems involve the recovery of structured signals from
a dimensionality reduced representation such as a random projection;
examples include sparse signals (compressive sensing) and low-rank
matrices (matrix completion). Given the wide range of different recovery
algorithms developed to date, it is natural to ask whether there exist
``universal'' algorithms for recovering ``structured'' signals from their
linear projections.  We recently answered this question in the affirmative
in the noise-free setting.  In this paper, we extend our results to the
case of noisy measurements.

\end{abstract}


\section{Introduction}\label{sec:intro}

Data compressors are ubiquitous in the digital world. They are built based on the premise  that text, images, videos, etc.\ all are  highly structured objects, and hence  exploiting those structures can dramatically reduce the number of bits required for their  storage.  In recent years,  a parallel trend has been developing for sampling analog signals. There too, the idea is that many analog signals of interest have some kind of structure that enables considerably lowering the sampling rate   from the  Shannon-Nyquist  rate.  

The  first structure that was extensively  studied in this context is {\em sparsity}. It has been observed that many natural signals have sparse representations in some  domain. 
The term compressed sensing (CS) refers to the process of undersampling a high-dimensional sparse signal through  linear measurements  and  recovering it  from those measurements using efficient algorithms \cite{Donoho1, CaRoTa06}. Low-rankedness \cite{ReFaPa10}, model-based compressed sensing \cite{RichModelbasedCS, eldar2010block, stojnic2009reconstruction, stojnic2009block, MaCeWi05}, and finite rate of innovation  \cite{VeMaBl02} are  examples of some other structures that have already  been explored in the literature.

While in the original source coding problem introduced by Shannon \cite{Shannon:48}, the assumption was that the source distribution is known both to the encoder and to the decoder, and hence is used in the code design, it was later shown that this information is not essential. In fact, \emph{universal} compression algorithms are able to code  stationary ergodic processes at their entropy rates, without knowing the source distribution \cite{cover}. In other words, there exists a family of compression codes that are able to  code any stationary ergodic process  at its entropy rate asymptotically \cite{cover}. The same result holds for universal lossy compression. 

One can ask similar questions for the problem of undersampling ``structured'' signals: How to define the class of  ``structured'' signals? Are there sampling and recovery algorithms for the reconstruction of ``structured'' signals from their linear measurements without having the knowledge of the underlying structure?  Does this ignorance incur a cost in the sampling rate?  

In algorithmic information theory, {\em Kolmogorov complexity}, introduced by Solomonoff \cite{Solomonoff}, Kolmogorov \cite{KolmogorovC}, and Chaitin \cite{chaitin66}, defines a universal notion of complexity for finite-alphabet sequences. Given a finite-alphabet sequence $x$, the Kolmogorov complexity of $x$, $K(x)$, is defined as the length of the shortest computer program that prints $x$ and halts.  In \cite{JalaliM:11}, extending the notion of Kolmogorov complexity to real-valued signals\footnote{These type of extensions are straightforward and have already been   explored in \cite{Staiger2002455}.} by their proper quantization, we addressed some of the above  questions.  We introduced the minimum complexity pursuit (MCP) algorithm for recovering ``structured'' signals  from their linear measurements.  We showed that finding the ``simplest'' solution satisfying the linear measurements recovers the signal using many fewer measurements than its ambient dimension. 

In this paper, we extend the results of \cite{JalaliM:11} to the case where the measurements are noisy. We first propose an updated version of MCP that takes into account that the measurements are a linear transformation of the signal plus Gaussian noise. We prove that the proposed algorithm is stable with respect to the noise and derive bounds on its reconstruction error in terms of the sampling rate and the variance of the noise.

The organization of this paper is as follows. Section \ref{sec:def} defines  the notation used throughout the paper. Section \ref{sec:kolm} defines Kolmogorov information dimension of a real-valued signal. Section \ref{sec:mcp} formally defines the MCP algorithm and reviews and extends some of the related results proved in \cite{JalaliM:11}. Section \ref{sec:contrib} considers the case of noisy measurements and proves that MCP is stable. Section \ref{sec:related} mentions some of the related work in the literature, and Section \ref{sec:conclusion} concludes the paper. Appendix A presents two useful lemmas used in the proofs. 


\section{Definitions}\label{sec:def}
\subsection{Notation}
Calligraphic letters such as $\Ac$ and $\Bc$ denote sets. For a set $\Ac$, $|\Ac|$ and $\Ac^c$ denote its size and its complement, respectively. 
For a sample space $\Omega$ and  event set $\Ac\subseteq \Omega$, $\ind_{\Ac}$ denotes the indicator function of the event $\Ac$. Boldfaced lower case letters denote vectors. For a vector $\x=(x_1,x_2,\ldots,x_n) \in \mathds{R}^n$, its $\ell_p$ and $\ell_{\infty}$ norms are defined as $\|x\|^p_p\triangleq\sum_{i=1}^n |x_i|^p$ and $\|x\|_{\infty}\triangleq\max_{i}|x_i|$, respectively. For integer $n$, let $I_n$ denote the $n\times n$ identity matrix.

For $x\in[0,1]$, let $((x)_1,(x)_2,\ldots$),  $(x)_i\in\{0,1\}$, denote the binary expansion of $x$, \ie  $x=\sum_{i=1}^{\infty}2^{-i}(x)_i$. The $m$-bit approximation  of $x$, $[x]_m$, is defined as  $[x]_m\triangleq\sum_{i=1}^{m}2^{-i}(x)_i$.
Similarly, for a vector $(x_1,\ldots,x_n)\in[0,1]^n$, $[x^n]_m\triangleq ([x_1]_m,\ldots,[x_n]_m)$.


\subsection{Kolmogorov complexity}\label{sec:kolm}
The Kolmogorov complexity of a finite-alphabet sequence $\x$ with respect to a universal Turing machine $\Uc$ is defined as the  length of the shortest program on $\Uc$ that prints $\x$ and halts.\footnote{See Chapter 14 of \cite{cover} for the exact definition of a universal computer, and more details on the definition of  Kolmogorov complexity.} Let $K(\x)$ denote the Kolmogorov complexity of  binary string $\x\in\{0,1\}^*\triangleq \cup_{n\geq 1}\{0,1\}^n$. 
\begin{definition}
For   real-valued  $\xv=(x_1,x_2,\ldots,x_n)\in [0,1]^n$, define the {\em Kolmogorov complexity of $\xv$ at  resolution $m$} as
\begin{align*}
K^{[\cdot]_m}(\x)  = K([x_1]_m,[x_2]_m,\ldots,[x_n]_m).
\end{align*}

\end{definition}

\begin{definition}
The {\em Kolmogorov information dimension} of  vector $(x_1, x_2, \ldots, x_n)\in[0,1]^n$ at resolution $m$ is defined as
\[
\kappa_{m,n} \triangleq \frac{K^{[\cdot]_{m}}(x_1,x_2, \ldots, x_n)}{m}.
\]
\end{definition} 

To clarify the above definition, we derive an upper bound for $\kappa_{m,n}$. 

\begin{lemma} \label{lemma:kappa-ub}
For  $(x_1, x_2, \ldots )\in [0,1]^{\infty}$ and any resolution sequence $\{m_n\}$, we have
\[
\limsup_{n \rightarrow \infty} \frac{\kappa_{m,n}}{   n} \leq 1.
\]
\end{lemma}
Therefore, by Lemma \ref{lemma:kappa-ub}, we call a  signal {\em compressible},  if $\limsup_{n \rightarrow \infty}n^{-1}\kappa_{m,n} < 1$.  As stated in the following proposition, Lemma \ref{lemma:kappa-ub}'s  upper bound on $\kappa_{m,n}$ is achievable. 

\begin{proposition}
Let $\{X_i\}_{i=1}^{\infty} \overset{iid}{\sim} {\rm Unif}[0,1]$. Then,
\[
\frac{1}{mn}K^{[\cdot]_m}(X_1, X_2, \ldots X_n) \rightarrow 1
\]
in probability.
\end{proposition}


\section{Minimum complexity pursuit}\label{sec:mcp}

Consider the problem of reconstructing a vector $x_o^n \in [0,1]^n$ from $d<n$ random linear measurements $y_o^d = Ax_o^n$. The MCP algorithm proposed in \cite{JalaliM:11} reconstructs  $x_o^n$ from its linear measurements $y_o^d$ by solving the following optimization problem:
\begin{eqnarray}
&& \min\limits_{x^n}\quad K^{[\cdot]_m}(x_1,\ldots,x_n)\nonumber \\
&&{\rm s.t.}\quad \ \   Ax^n = y_o^d.\label{eq:alg}
\end{eqnarray}

Let  the elements of $A\in\mathds{R}^{d\times n}$,  $A_{ij}$, be i.i.d.\ $\Nc(0,1)$.\footnote{Note that in \cite{JalaliM:11} we had assumed that $A_{i,j}\sim\Nc(0,1/d)$.}  Let $\xh_o^n=\xh_o^n(y_o^d,A)$ denote the output of \eqref{eq:alg} to  inputs $y_o^d=Ax_{o}^n$ and $A$. Theorem \ref{thm:1} stated below is a generalization of Theorem 2 proved in \cite{JalaliM:11}.

\begin{theorem}\label{thm:1}
Assume that $x_o=(x_{o,1},x_{o,2},\ldots)\in[0,1]^{\infty}$. For integers $m$ and $n$, let $\kappa_{m,n}$ denote the Kolmogorov information dimension of $x_o^n$ at resolution $m$.  Then, for any $\tau_n <1$ and $t>0$, we have
\begin{align*}
&\P \left(\|x_{o}^n-\xh_{o}^n\|_2>  \frac{ \sqrt{nd^{-1}+t + 1} +1}{\tau_n} \sqrt{n 2^{-2m+2}}\right) \nonumber \\
&  \leq 2^{2\kappa_{m,n} m} {\rm e}^{\frac{d}{2} (1- \tau_n^2  +2 \log \tau_n )} + {\rm e}^{- \frac{d}{2} t^2}.
\end{align*}

\end{theorem}

Theorem \ref{thm:1} can be proved following the steps used in the proof of Theorem 2 in \cite{JalaliM:11}. To interpret this theorem, in the following we consider several interesting corollaries that follow from Theorem \ref{thm:1}. Note that in all of the results, the logarithms are to the base of Euler's number $e$.

\begin{corollary}\label{col:1}
Assume that $x_o=(x_{o,1},x_{o,2},\ldots)\in[0,1]^{\infty}$ and $m=m_n = \lceil \log n \rceil$. Let $\kappa_n\triangleq \kappa_{m_n,n}$. Then if $d_n = \lceil \kappa_n \log n\rceil$, for any $\e>0$,  we have
$\P\left(\|x_{o}^n-\xh_{o}^n\|_2> \e \right) \rightarrow 0$,
as $n \rightarrow \infty$.
\end{corollary}
{\em Proof:} For $m=m_n= \lceil \log n \rceil$ and $d_n =\lceil\kappa_n \log n \rceil$,
\begin{eqnarray}
\lefteqn{(\sqrt{nd^{-1}+t + 1} +1) \sqrt{n 2^{-2m_n+2}}} \nonumber\\
& \leq &2\left(\sqrt{\lceil \kappa_n \log n\rceil^{-1}+(t + 1)n^{-1}} +\sqrt{n^{-1}}\right). 
\end{eqnarray}
Hence, fixing $t>0$ and setting $\tau_n =\tau= 0.1$, for any $\e>0$ and large enough values of $n$ we have
\[
{(\sqrt{nd^{-1}+t + 1} +1) \sqrt{n 2^{-2m+2}}\over \tau_n} \leq \e.
\]
Therefore, for $n$ large enough,
\begin{eqnarray}
\lefteqn{\P\left(\|x_{o}^n-\xh_{o}^n\|_2^2> \e \right) } \nonumber \\
&\leq& 2^{2\kappa_{n}  \log n} {\rm e}^{\frac{ \kappa_n\log n }{2} (1- \tau^2  +2 \log \tau )} + {\rm e}^{- \frac{d}{2} t^2} \nonumber\\
&\leq& e^{1.4\kappa_{n}\log n} {\rm e}^{-1.7\kappa_n\log n} + {\rm e}^{- \frac{d}{2} t^2},
\end{eqnarray}
which shows that as $n\to\infty$, $\P(\|x_{o}^n-\xh_{o}^n\|_2^2> \e )\to 0$.

 $\hfill \Box$

According to Corollary \ref{col:1},   if the complexity of the signal is less than $\kappa$, then the number of linear measurements needed for asymptotically perfect recovery is, roughly speaking, at the order of $\kappa \log n$. In other words, the number of measurements is proportional to the complexity of the signal and only logarithmically proportional to its ambient dimension.

\begin{corollary}
Assume that $x_o=(x_{o,1},x_{o,2},\ldots)\in[0,1]^{\infty}$ and $m=m_n = \lceil \log n \rceil$. Let $\kappa_n\triangleq \kappa_{m_n,n}$.  Then, if $d=d_n = \lceil 3 \kappa_n \rceil $,  we have
\[
\P\left(\frac{1}{\sqrt{n}}\|x_{o}^n-\xh_{o}^n\|_2> \e \right) \rightarrow 0,
\]
as $n \rightarrow \infty$,  for any $\e>0$.
\end{corollary}
{\em Proof:} Setting $\tau_n = n^{-0.5}$, $m=m_n = \lceil \log n \rceil$, and $d=d_n=\lceil 3 \kappa_n \rceil$ in Theorem \ref{thm:1}, it follows that 
\begin{align*}
\P& \left({1\over {\sqrt{n}}}\|x_{o}^n-\xh_{o}^n\|_2> 2\sqrt{d_n^{-1}+(t+1)n^{-1}} +2\sqrt{n^{-1}} \right) \nonumber \\
&\leq 2^{2\kappa_{n}\log n} {\rm e}^{1.5\kappa_n (1- n^{-1} -\log n )} + {\rm e}^{- \frac{d}{2} t^2}  \nonumber\\
&= e^{-(1.5-2\log 2)\kappa_{n}\log n + \kappa_n(1.5- 1.5n^{-1})} + {\rm e}^{- \frac{d}{2} t^2}. 
\end{align*}
Since $1.5-2\log 2>0$, for any  $\e>0$ and $n$ large enough, we have 
\[
2\sqrt{d_n^{-1}+(t+1)n^{-1}} +2\sqrt{n^{-1}}<\e.
\]
 It follows that $\P(\frac{1}{\sqrt{n}}\|x_{o}^n-\xh_{o}^n\|_2> \e) \to 0$, as $n\to\infty$.

$\hfill\Box$

In other words, if we are interested in the normalized mean square error, or per element squared distance, then $3 \kappa_n$ measurements are sufficient.


\section{Stability analysis of MCP}\label{sec:contrib}

In the previous section we considered the case where the signal is exactly of low complexity and the measurements are also noise-free. In this section, we extend the results to noisy measurements, where $y_o^d = Ax_o^n + w^d$, with $w^d \sim \Nc(0, \sigma^2 I_d)$. Assuming that the complexity of the signal is known at the reconstruction stage, we consider the following reconstruction algorithm:
\begin{eqnarray}
&& \arg\min_{x^n}\quad \|Ax^n-y_o^d\|^2_2, \nonumber \\
&&{\rm s.t.}\quad \quad \quad  K^{[\cdot]_m}(x^n) \leq \kappa_{m,n} m.\label{eq:alg_noisy}
\end{eqnarray}
Note that $\kappa_{m,n} m$ is an upper bound on the Kolmogorov complexity of $x_o$ at resolution $m$. The major issue of this section is to calculate the number of measurements required for  robust recovery in noise. 

\begin{theorem}\label{thm:2}
Assume that $x_o=(x_{o,1},x_{o,2},\ldots)\in[0,1]^{\infty}$. For integers $m$ and $n$, let $\kappa_{m,n}$ denote the information dimension of $x_o^n$ at resolution $m$.   If  $m=m_n=\lceil\log n\rceil $ and $d=8r\kappa_{m,n}m$, where $r>1$, then for any $\e >0$, we have
\begin{align}
\P\left(\|x_{o}^n-\xh_{o}^n\|_2^2 >  {(2\kappa_{m,n}m)\sigma^2 \over \rho d} \right) \to 0,
\end{align}
as $n\to\infty$, where $\rho \triangleq (1-\sqrt{r^{-1}})^2/2$.
\end{theorem}

According Theorem \ref{thm:2}, as long as $d > 8r \kappa_n \log n$ the algorithm is stable in the sense that the reconstruction error is proportional to the variance of the input noise. By increasing the number of measurements one may reduce the reconstruction error. 

{\em Proof:} Since by definition $K^{[\cdot]_m}(x_o^n) = k_{m,n} m_n$, $x_o^n$ is also a feasible point in \eqref{eq:alg_noisy}. But, by assumption, $\xh_o^n$ is the solution of \eqref{eq:alg_noisy}. Therefore,
\begin{align}
\|A\xh_o^n-y_o^d\|^2_2&\leq \|A x_o^n-y_o^d\|^2_2\nonumber\\
&=\|A x_o^n-Ax_o^n-w^d\|^2_2=\|w^d\|^2_2.\label{eq:basic-ineq}
\end{align}
Expanding $\|A\xh_o^n-y_o^d\|^2_2=\|A\xh_o^n-Ax_o^n-w^{d}\|^2_2$ in \eqref{eq:basic-ineq}, it follows that 
\begin{align}
&\|A(\xh_o^n-x_o^n)\|^2_2 + \|w^d\|^2_2  -2(w^d)^TA(\xh_o^n-x_o^n) \leq \|w^d\|^2_2.\label{eq:basic-ineq-expanded}
\end{align}
Canceling $ \|w^d\|^2_2$ from both sides of \eqref{eq:basic-ineq-expanded}, we obtain
\begin{align*}
\|A(\xh_o^n-x_o^n)\|^2_2 &\leq   2(w^d)^TA(\xh_o^n-x_o^n)\nonumber\\
&\leq   2\left|(w^d)^TA(\xh_o^n-x_o^n)\right|.
\end{align*}

Let $e_m^n\triangleq x_o^n-[x_o^n]_m$ and $\eh_m^n\triangleq\xh_o^n-[\xh_o^n]_m$ denote the quantization errors of the original and the reconstructed signals, respectively. Using these definitions, and the Cauchy-Schwartz inequality, we find a lower bound for $\|A(\xh_o^n-x_o^n)\|^2_2$ as 
\begin{align}
\|A&(\xh_o^n-x_o^n)\|^2_2 \nonumber\\
&= \|A([\xh_o^n]_m+\eh_m^n - [x_o^n]_m-e_m^n)\|^2_2 \nonumber\\
 &= \|A([\xh_o^n]_m- [x_o^n]_m)+A(\eh_m^n -e_m^n)\|^2_2 \nonumber\\
&\geq \|A([\xh_o^n]_m- [x_o^n]_m)\|_2^2 \nonumber\\
&\quad-2\left| (\eh_m^n -e_m^n)^TA^TA\left([\xh_o^n]_m- [x_o^n]_m\right)\right|\nonumber\\
&\geq \|A([\xh_o^n]_m- [x_o^n]_m)\|_2^2 \nonumber\\
&\quad-2\left\|A(\eh_m^n -e_m^n)\right\|_2\left\|A\left([\xh_o^n]_m- [x_o^n]_m\right)\right\|_2.\label{eq:lower}
\end{align}

On the other hand, again using our definitions plus the Cauchy-Schwartz  inequality, we find an upper bound on $|(w^d)^TA(\xh_o^n-x_o^n)|$ as 
\begin{align}
&\left|(w^d)^TA(\xh_o^n-x_o^n)\right|\nonumber\\
&=\left|([\xh_o^n]_m - [x_o^n]_m+\eh_m^n-e_m^n)^TA^Tw^d\right|\nonumber\\
&\leq \left|([\xh_o^n]_m - [x_o^n]_m)^TA^Tw^d\right|+\left|(\eh_m^n-e_m^n)^TA^Tw^d\right|\nonumber\\
&\leq \left|([\xh_o^n]_m - [x_o^n]_m)^TA^Tw^d\right|+\|\eh_m^n-e_m^n\|_2\|A^Tw^d\|_2.\label{eq:upper}
\end{align}

For any  $x\in[0,1]$, $0\leq x-[x]_m < 2^{-m}$. Therefore, 
\begin{align}
\|\eh_m^n-e_m^n\|_2\leq \sqrt{n2^{-2m+2}}.\label{eq:ell2-error}
\end{align}

Let set $\Sc$ be the set of all vectors of length $n$ that can be written as the difference of two vectors with complexity less than $k_{m,n}m$; that is,
\[
\Sc=\left\{h_1^n-h_2^n:  K(h_1^n)\leq \kappa_{m,n}m, \,  K(h_2^n)\leq \kappa_{m,n}m\right\}.
\]
Note that $|\Sc|\leq 2^{2\kappa_{m,n}m }$. Define the event $\Ec_1$ as 
\[
\Ec_1\triangleq \{\forall\; h^n\in\Sc: \|(w^d)^TA h^n\|\leq  t_1 \|h^n\|_2\}.
\]

For any fixed $h^n$, $Ah^n$ is an i.i.d.\ zero-mean Gaussian vector of length $d$ and variance $\|h^n\|_2^2$. Assuming that $\|h^n\|_2=1$ and applying Lemma \ref{lemma:gaussian-vectors}, we obtain
\begin{align}
\P&\left(\left|(w^d)^TAh^n\right|\geq t_1\right)=\P\left(\|w^d\|_2 G\geq t_1\right)\nonumber\\
&=\P\left(\left\|w^d\right\|_2 G\geq t_1, \|w^d\|_2\geq \sqrt{d}\sigma(1+t_2) \right)\nonumber\\
&\quad+\P\left(\left\|w^d\right\|_2 G\geq t_1, \|w^d\|_2< \sqrt{d}\sigma(1+t_2) \right)\nonumber\\
&\leq \P\left(\|w^d\|_2\geq \sqrt{d}\sigma(1+t_2) \right)\nonumber\\
&\quad+\P\left(G\geq t_1(\sqrt{d}\sigma(1+t_2))^{-1} \right)\nonumber\\
&\leq e^{-dt_2^2/2}+e^{-{t_1^2\over2\sigma^2d(1+t_2)}}.
\end{align}
Hence, by the union bound and the fact that  $|\Sc|\leq 2^{2\kappa_{m,n}m}$ \cite{cover}, we have 
\begin{align} 
\P(\Ec_1^c) \leq  2^{2\kappa_{m,n}m}\left(e^{-dt_2^2/2}+e^{-{t_1^2\over2\sigma^2d(1+t_2)}}\right).\label{eq:pe1}
\end{align}

Note that
\begin{align*}
\left\|A(\eh_m^n -e_m^n)\right\|_2\leq \sigma_{\max}(A)\|e_m^n -e_m^n\|_2.
\end{align*}
For $t_3>0$, define the event $\Ec_2$ as
\begin{align*}
\Ec_2^{(n)}\triangleq \left\{\sigma_{max}(A)   < (1+t_3)\sqrt{d}+\sqrt{n} \right\}.
\end{align*}
It can be proved that \cite{CaTa05}
\begin{align}
\P\left(\Ec_2^{(n),c}\right) \leq {\rm e}^{-d t_3^2/2}. \label{eq:pe2}
\end{align}
But if  $\sigma_{\max}(A)   < (1+t_3)\sqrt{d}+\sqrt{n} $, then from \eqref{eq:ell2-error}
\begin{align*}
\left\|A(\eh_m^n -e_m^n)\right\|_2\leq  \left(1+(1+t_3)\sqrt{d \over n}\right) 2^{-m+1}n.
\end{align*}

Define the event $\Ec_3^{(n)}$ as $\Ec_3^{(n)}\triangleq\{ \forall \; h^n\in\Sc:  \|A h^n\|^2_2 > (1-t_4) d \|h^n\|^2_2 \}$. By the union bound and Lemma \ref{lemma:chi}, it follows that 
\begin{align}
&\P\left(\Ec_3^{(n),c}\right) \leq 2^{2\kappa_{m,n}m} \rm{\rm e}^{\frac{d}{2}(t_4+  \log (1-t_4)) }.\label{eq:pe3}
\end{align}

Define  the  event $\Ec_4^{(n)}$ as
\[
\Ec_4^{(n)}\triangleq\{ \forall \; h^n\in\Sc;  \|A h^n\|^2_2 <{(1+t_5)  d} \|h^n\|^2_2 \},
\]
Again by the union bound and Lemma \ref{lemma:chi}, it follows that 
\begin{align}
&\P\left(\Ec_4^{(n),c}\right) \leq 2^{2\kappa_{m,n}m} \rm{\rm e}^{-\frac{d}{2}(t_5-  \log (1-t_5)) }.\label{eq:pe4}
\end{align}

Finally, for $t_6>0$, define
\[
\Ec_5^{(n)}\triangleq\{\|A^Tw^d\|_2^2 \leq nd(1+t_6)  \}.
\]
Given $w^d$, $A^Tw^d$ is an $n$ dimensional i.i.d.\ Gaussian random vector with variance $\|w^d\|_2^2$. Hence, by Lemma \ref{lemma:chi},
\begin{align*}
\P&\left(\|A^Tw^d\|_2^2 \geq n\gamma^2(1+t_7) \left|\|w^d\|_2^2=\gamma^2 \right.\right)\nonumber\\
&\leq  {\rm e} ^{-\frac{n}{2}(t_7 - \log(1+ t_7))}.
\end{align*}
 On the other hand, again by Lemma \ref{lemma:chi},
\begin{align*}  
\P(\|w^d\|_2^2<d(1-t_8)) \leq  {\rm e} ^{\frac{d}{2}(t_8 + \log(1- t_8))}.
\end{align*}
Choosing $t_6,t_7,t_8>0$ such that $t_6<t_7$ and $1+t_6=(1-t_8)(1+t_7)$, it follows that
\begin{align}
\P&\left(\|A^Tw^d\|_2^2 \geq nd(1+t_6) \right)\nonumber\\
&=\P\left(\|A^Tw^d\|_2^2 \geq nd(1+t_6),  \|w^d\|_2^2>d(1-t_8) \right)\nonumber\\
&\;\;+\P\left(\|A^Tw^d\|_2^2 \geq nd(1+t_6) , \|w^d\|_2^2<d(1-t_8) \right)\nonumber\\
&\leq   {\rm e} ^{-\frac{n}{2}(t_7 - \log(1+ t_7))} +  {\rm e} ^{\frac{d}{2}(t_8 + \log(1- t_8))}.\label{eq:pe5}
\end{align}

Combining \eqref{eq:lower} and \eqref{eq:upper} and the upper and lower bounds derived for the corresponding terms of  \eqref{eq:lower} and \eqref{eq:upper}, and choosing $t_1=2\sigma\sqrt{d(1+t_2)(2\kappa_{m,n}m)}$, with probability $\P(\Ec_1\cap \Ec_2\cap\Ec_3\cap\Ec_4\cap\Ec_5), $ the following  inequality holds:
\begin{align}
&(1-t_4)\sqrt{d}\|\Delta_m\|_2^2\nonumber\\
&-2\left(\sqrt{1+t_5}2^{-m+1}\sqrt{n}((1+t_3)\sqrt{d}+\sqrt{n}))\right)\|\Delta_m\|_2\nonumber\\
&-2\left(\sigma \sqrt{1+t_2}\sqrt{2\kappa_{m,n}m}\right)\|\Delta_m\|_2\nonumber\\
&-2^{-m+1}\sqrt{1+t_6}n \leq 0.\label{eq:main-2nd-order-ineq}
\end{align}
Inequality \eqref{eq:main-2nd-order-ineq} involves a quadratic equation of $\|\Delta_m\|_2$. Finding the roots of this quadratic equation,  using $\sqrt{1+x}\leq 1+x/2$, and replacing $m=\lceil\log n\rceil$, we obtain
\begin{align}
\|\Delta_m\|_2\leq& \;\sigma \gamma_3 \sqrt{(2\kappa_{m,n} \log n) d^{-1}} \nonumber\\
&+(\gamma_1\sqrt{n^{-1}}+\gamma_2{ \sqrt{d^{-1}}})+\sqrt{d^{-1}}\gamma_4,
\end{align} 
where $\gamma_1=\sqrt{1+t_5}(1+t_3)(1-t_4)^{-1}$, $\g_2=\sqrt{1+t_5}(1-t_4)^{-1}$, $\g_3=\sqrt{1+t_2}(1-t_4)^{-1}$ and $\g_4=\sqrt{1+t_6}(1-t_4)^{-1}$. On the other hand, by the union bound,
\begin{align}
&\P\left((\Ec_1\cap \Ec_2\cap\Ec_3\cap\Ec_4\cap\Ec_5)^c\right)= \P(\Ec_1^c\cup \Ec_2^c\cup\Ec_3^c\cup\Ec^c_4\cup\Ec^c_5)\nonumber\\
&\leq \P(\Ec_1^c)+\P(\Ec_2^c)+\P(\Ec_3^c)+\P(\Ec_4^c)+\P(\Ec_5^c).\label{eq:pe-final}
\end{align}
Given $d=8r\kappa_{m,n}m$, choosing $t_2=t_4=1/\sqrt{r}$ and fixing $t_1$, $t_3$, $t_5,\ldots,t_8$ at appropriate  fixed small numbers,   \eqref{eq:pe1}, \eqref{eq:pe2}, \eqref{eq:pe3}, \eqref{eq:pe4} and \eqref{eq:pe5} guarantee that \eqref{eq:pe-final} goes to zero, as $n\to\infty$. Moreover, for chosen parameters, $\gamma_3<\sqrt{2}/(1-\sqrt{r^{-1}})$. Finally, for any $\e>0$, for $n$ large enough, $(\gamma_1\sqrt{n^{-1}}+\gamma_2{ \sqrt{d^{-1}}})+\sqrt{d^{-1}}\gamma_4 <\e$. This concludes the proof.$\hfill\Box$


\section{Related work}\label{sec:related}

The MCP algorithm proposed in \cite{JalaliM:11} is mainly inspired by \cite{DonohoKS2002} and \cite{DoKaMe06}.  Consider the universal denoising problem where $\boldsymbol{\theta}$ is corrupted by additive white Gaussian noise as $X^n=\boldsymbol{\theta}+Z^n$. The denoiser's goal is to recover  $\boldsymbol{\theta}$ from the  noisy observation  $X^n$.  The \textit{minimum Kolmogorov complexity estimation} (MKCE) approach proposed in \cite{DonohoKS2002} suggests a denoiser that looks for the  sequence $\hat{\boldsymbol{\theta}}$ with minimum Kolmogorov complexity among all the vectors that are within some distance of the observation vector $X^n$.   \cite{DonohoKS2002} shows that if $\t_i$ are i.i.d., then under certain conditions, the average marginal conditional distribution of $\hat{\theta}_i$ given $X_i$  tends to the actual posterior distribution of $\theta_1$ given $X_1$.

In \cite{DonohoKS2002}, the authors consider the problem of recovering a  low-complexity sequence from its linear  measurements. Let
$\Sc(k_0)\triangleq\{x^n\in [0,1]^n: K(x^n)\leq k_0\}$.
Consider measuring $x_o^n\in\Sc(k_0)$ using a $d\times n$ binary matrix $A$. Let  $y_o^d=Ax_o^n$.  To recover $x_o^n$ from measurements $y_o^d$,  \cite{DonohoKS2002} suggests finding  $\xh^n$ as ${\xh}^n(y_o^d,A) \triangleq \argmin_{x^n:\;y_o^d=Ax^n} K(x^n)$, and proves that  if $d\geq 2k_0$, then this algorithm is able to find $x_o^n$ with high probability. Clearly assuming that a real-valued sequence has a low complexity is very restrictive, and hence $\Sc(k_0)$ does not include any of the classes that has been studied in CS literature. For instance most of the one sparse signals have infinite Kolmogorov complexity, and hence the result of \cite{DonohoKS2002} does not imply useful information. 

In a recent and independent work, \cite{BaronD:11} and \cite{BaDu12} consider a scheme similar to MCP. For a stationary and ergodic source, they propose an algorithm to approximate MCP. While the empirical results are promising, no theoretical guarantees are provided on either the performance of MCP or their final algorithm.

The notion of sparsity has already been generalized in the literature in several different directions  \cite{RichModelbasedCS, ChRePaWi10, VeMaBl02, ReFaPa10}. More recently,  \cite{ChRePaWi10} introduced the  class of simple functions and atomic norm as a framework that unifies some of the above observations and extends them  to some other signal classes.  While all these models can be considered as subclasses of the general model  considered in this paper,  it is worth noting that even though the recovery approach proposed here is universal, given the incomputibility of Kolmogorov complexity,  it is not useful for practical purposes. Finding practical algorithms with provable performance guarantees is left for future research.  

In this paper, we have focused on  deterministic signal models.  For the case of  random signals, \cite{WuVe10} considers the problem of recovering a memoryless process from its undersampled  linear measurements and establishes a connection between the required  number of measurements  and the Renyi entropy of the source. Also, our work is in the same spirit as the minimum entropy  decoder proposed by Csiszar in \cite{Csiszar82}, which is a universal  decoder, for reconstructing an i.i.d.\ signal from its linear measurements.


\section{Conclusion}\label{sec:conclusion}

In this paper, we have  considered the problem of recovering structured signals from their random linear measurements. We have investigated the minimum complexity pursuit (MCP) scheme. Our results confirm that if the Kolmogorov complexity of the signal is upper bounded by $\kappa$, then MCP recovers the signal accurately from $O(\kappa \log n)$ random linear measurements, which is much smaller than the ambient dimension. In this paper, we have specifically proved that MCP is stable, such that the $\ell_2$-norm of the reconstruction error is proportional to the standard deviation of the noise.  


\renewcommand{\theequation}{A-\arabic{equation}}
\setcounter{equation}{0}  

\section*{Appendix A} \label{sec:proofs}

The following two lemmas are  frequently used in our proofs. 

\begin{lemma}[$\chi$-square concentration]\label{lemma:chi}
Fix $\tau>0$, and let $Z_i\sim\Nc(0,1)$, $i=1,2,\ldots,d$. Then,
$\P( \sum_{i=1}^d  Z_i^2 <d(1- \tau))  \leq {\rm e} ^{\frac{d}{2}(\tau + \log(1- \tau))}$, 
and 
\[
\P( \sum_{i=1}^d  Z_i^2 > d(1+\tau))  \leq {\rm e} ^{-\frac{d}{2}(\tau - \log(1+ \tau))}.
\]
\end{lemma}

The proof of Lemma \ref{lemma:chi} is presented in \cite{JalaliM:11}.

\begin{lemma}\label{lemma:gaussian-vectors}
Let $X^n$ and $Y^n$ denote two independent Gaussian random vectors with i.i.d.\ elements. Further assume that for $i=1,\ldots,n$,  $X_i\sim\Nc(0,1)$ and  $Y_i\sim\Nc(0,1)$. Then the distribution of  $(X^n)^Ty^n=\sum_{i=1}^nX_iY_i$ is the same as the distribution of $\|X^n\|_2G$, where $G\sim\Nc(0,1)$ is independent of $\|X^n\|_2$.
\end{lemma}

{\em Proof:} We need to show that  $(X^n)^TY^n/\|X^n\|_2$ is distributed as $\Nc(0,1)$ and is independent of $\|X^n\|_2$. To prove the first claim, note that
\begin{align}
{(X^n)^TY^n\over \|X^n\|_2}&=\sum_{i=1}^n{X_i\over \|X^n\|_2}Y_i.
\end{align}
On the other hand, given $X^n/\|X^n\|_2=a^n$, 
\[
\sum_{i=1}^n{X_i\over \|X^n\|_2}Y_i\sim \Nc(0,1),
\]
because $\sum_{i=1}^na_i^2=1$. Therefore, since the distribution of $\sum_{i=1}^n{X_i\over \|X^n\|_2}Y_i$ given $X^n/\|X^n\|_2=a^n$ is independent of $a^n$,  
\[\sum_{i=1}^n{X_i\over \|X^n\|_2}Y_i\sim \Nc(0,1).\]

To prove the independence, note that $X^n/\|X^n\|_2$ and $Y^n$ are both independent of $\|X^n\|_2$.

$\hfill\Box$

\bibliographystyle{unsrt}
\bibliography{myrefs}

\begin{thebibliography}{10}

\bibitem{Donoho1}
D.~L. Donoho.
\newblock Compressed sensing.
\newblock {\em IEEE Trans. Info. Theory}, 52(4):489--509, April 2006.

\bibitem{CaRoTa06}
E.~Cand\`es, J.~Romberg, , and T.~Tao.
\newblock Robust uncertainty principles: Exact signal reconstruction from
  highly incomplete frequency information.
\newblock {\em IEEE Trans. Info. Theory}, 52(2):489--509, February 2006.

\bibitem{ReFaPa10}
B.~Recht, M.~Fazel, and P.~A. Parrilo.
\newblock Guaranteed minimum rank solutions to linear matrix equations via
  nuclear norm minimization.
\newblock {\em {SIAM} Review}, 52(3):471--501, April 2010.

\bibitem{RichModelbasedCS}
R.~G. Baraniuk, V.~Cevher, M.~F. Duarte, and C.~Hegde.
\newblock Model-based compressive sensing.
\newblock {\em IEEE Trans. Info. Theory}, 56(4):1982 --2001, April 2010.

\bibitem{eldar2010block}
Y.~C. Eldar, P.~Kuppinger, and H.~Bolcskei.
\newblock Block-sparse signals: Uncertainty relations and efficient recovery.
\newblock {\em IEEE Trans. on Sig. Proc.}, 58(6):3042--3054, 2010.

\bibitem{stojnic2009reconstruction}
M.~Stojnic, F.~Parvaresh, and B.~Hassibi.
\newblock On the reconstruction of block-sparse signals with an optimal number
  of measurements.
\newblock {\em IEEE Trans. on Sig. Proc.}, 57(8):3075--3085, 2009.

\bibitem{stojnic2009block}
M.~Stojnic.
\newblock Block-length dependent thresholds in block-sparse compressed sensing.
\newblock {\em Arxiv preprint arXiv:0907.3679}, 2009.

\bibitem{MaCeWi05}
D.~Malioutov, M.~Cetin, and A.S. Willsky.
\newblock A sparse signal reconstruction perspective for source localization
  with sensor arrays.
\newblock {\em IEEE Trans. on Sig. Proc.}, 53(8):3010--3022, August 2005.

\bibitem{VeMaBl02}
M.~Vetterli, P.~Marziliano, and T.~Blu.
\newblock Sampling signals with finite rate of innovation.
\newblock {\em IEEE Trans. on Sig. Proc.}, 50(6):1417--1428, June 2002.

\bibitem{Shannon:48}
C.~E. Shannon.
\newblock A mathematical theory of communication: Parts {I and II}.
\newblock {\em Bell Syst. Tech. J.}, 27:379--423 and 623--656, 1948.

\bibitem{cover}
T.~Cover and J.~Thomas.
\newblock {\em Elements of Information Theory}.
\newblock Wiley, New York, 2nd edition, 2006.

\bibitem{Solomonoff}
R.~J. Solomonoff.
\newblock A formal theory of inductive inference.
\newblock {\em Inform. Contr.}, 7:224--254, 1964.

\bibitem{KolmogorovC}
A.~N. Kolmogorov.
\newblock Logical basis for information theory and probability theory.
\newblock {\em IEEE Trans. Info. Theory}, 14:662--664, 1968.

\bibitem{chaitin66}
G.~J. Chaitin.
\newblock On the length of program for computing binary sequences.
\newblock {\em J. Assoc. Comput. Mach.}, 13:547 -- 569, 1966.

\bibitem{JalaliM:11}
S.~Jalali and A.~Maleki.
\newblock Minimum complexity pursuit.
\newblock In {\em Proc. 49th Allerton Conference on Communication, Control, and
  Computation}, Monticello, IL, Sep. 2011.

\bibitem{Staiger2002455}
L.~Staiger.
\newblock The {Kolmogorov} complexity of real numbers.
\newblock {\em Theoretical Computer Science}, 284(2):455 -- 466, 2002.

\bibitem{CaTa05}
E.~Cand\`es, J.~Romberg, , and T.~Tao.
\newblock Decoding by linear programming.
\newblock {\em IEEE Trans. Info. Theory}, 51(12):4203--4215, Dec. 2005.

\bibitem{DonohoKS2002}
D.~L. Donoho.
\newblock Kolmogorov sampler.
\newblock {\em Preprint}, 2002.

\bibitem{DoKaMe06}
D.~L. Donoho, H.~Kakavand, and J.~Mammen.
\newblock The simplest solution to an underdetermined system of linear
  equations.
\newblock In {\em Proc. IEEE Int. Symp. Info. Theory}, July 2006.

\bibitem{BaronD:11}
D.~Baron and M.~Duarte.
\newblock Universal {MAP} estimation in compressed sensing.
\newblock In {\em Proc. of 49th Allerton Conference on Communication, Control,
  and Computation}, Monticello, IL, Sep. 2011.

\bibitem{BaDu12}
D.~Baron and M.~Duarte.
\newblock Signal recovery in compressed sensing via universal priors.
\newblock {\em Arxiv preprint arXiv:1204.2611}, 2012.

\bibitem{ChRePaWi10}
V.~Chandrasekaran, B.~Recht, P.~A. Parrilo, and A.~Willsky.
\newblock The convex geometry of linear inverse problems.
\newblock {\em Preprint}, 2010.

\bibitem{WuVe10}
Y.~Wu and S.~Verd\'u.
\newblock Renyi information dimension: Fundamental limits of almost lossless
  analog compression.
\newblock {\em IEEE Trans. Info. Theory}, 56(8):3721--3748, Aug. 2010.

\bibitem{Csiszar82}
I.~Csiszar.
\newblock Linear codes for sources and source networks: Error exponents,
  universal coding.
\newblock {\em IEEE Trans. Info. Theory}, 28:585--592, 1982.

\end{thebibliography}

\end{document}